\documentclass[twocolumn,pre,amsmath,amssymb,superscriptaddress,english]{revtex4}
\usepackage{array}
\usepackage{tabularx}
\usepackage{dcolumn}
\usepackage{bm}
\usepackage{graphicx}
\usepackage{amsmath}
\usepackage{amssymb}
\usepackage{epsfig}
\usepackage{graphicx}
\usepackage{float}
\usepackage[usenames,dvipsnames]{color}
\usepackage{mathtools}
\usepackage{marginnote}
\usepackage{etoolbox}
\usepackage{lipsum}
\usepackage{color}
\usepackage{hyperref}

\UseRawInputEncoding

\usepackage{ae}
\usepackage{aecompl}
\usepackage{mathrsfs}

\usepackage[T1,OT1]{fontenc} 

\DeclareTextCommand{\DJ}{OT1}{%
  \raisebox{-0.1ex}{\scalebox{0.75}[1.4]{--}}\kern-.4em D%
}

\newcommand{\ie}{{\it i.e.}}

\begin{document}
\title{State-Resolved Integral of First-Passage Times for Multi-Site Polymer Adsorption}
\author{Yifan Huang}
\address{Key Laboratory of Quantum Materials and Devices of Ministry of Education, School of Physics, Southeast University, Nanjng 211189, China}
\author{Qiyun Tang}
\altaffiliation{qtang@seu.edu.cn }
\address{Key Laboratory of Quantum Materials and Devices of Ministry of Education, School of Physics, Southeast University, Nanjng 211189, China}
\begin{abstract}
Understanding the interfacial structure of multi-site polymer adsorption is critical for the rational design of functional nanomaterials. However, the distribution of chains attached via one, two, or more anchor points has remained inaccessible to experiments and conventional simulations. Experiments typically measure ensemble averages such as total adsorbed mass, while molecular dynamics simulations are limited to timescales far shorter than the relevant adsorption processes. To address this challenge, we introduce the State-Resolved Integral of First-Passage Times (SR-IFS) method. This approach decouples fast intra-layer conformational adjustments from the slow kinetics of external chain exchange, enabling quantitative prediction of the time-dependent distribution of attachment states ($p_1$, $p_2$, $p_3$) within a multi-site adsorbed layer. Using 3-arm star-like polymers as a model system, we show that the interfacial state distribution evolves from an initial prevalence of three-point attachments to a more heterogeneous mixture over long time scales, and that the equilibrium distribution can be systematically tuned by adjusting the monomer binding energy. The SR-IFS method provides access to this state-resolved information, offering a connection between microscopic kinetics and macroscopic interfacial properties.

\end{abstract}
\maketitle

\section{Introduction}

Integrating functional polymers with solid surfaces is fundamental to fabricating advanced nanomaterials \cite{Rossner2020planetsatellite, Scott2008protein, Gong2019simulating, Wisniewska2016impact, Sassolas2012immobilization, Lvov2016halloysite}, including nanopore sensors \cite{Awasthi2020polymer, Sze2017single, Vreeker2025nanoporefunctionalized, Wang2024chronopotentiometric, Yusko2017realtime, Hu2024translocation}, drug delivery carriers \cite{Lucky2015nanoparticles, Ou2020nanodrug, Qi2018block, Lu2019nonsacrificial}, surface coatings \cite{Choueiri2016surface, Duan2020solutionprocessed, He2021continuous, Lee2020electrolyte, Nie2008patterning, Wang2018triboelectrificationinduced, Yu2021solventevaporation}, and nanocomposite scaffolds \cite{Zhao2017bioinspired, Li2022design, Jiang2022radially}. In many such systems, polymers carry multiple binding sites that can attach simultaneously to the surface \cite{Tang2020prediction, Rossner2017uniform, Li2022design, Thoms2024simple, Zhou2025adsorption, Li2025model}. This multi-site adsorption creates structurally complex interfacial layers where individual chains may be attached via one, two, or more anchor points \cite{Tang2020prediction, Rossner2017uniform, Li2025model}. The relative fractions of these different attachment states, \ie, the interfacial state distribution, determine the layer's density, mechanical properties, permeability, and overall functionality \cite{Welch2015trains, Taylor2025dependence, Rossner2017uniform}.

Despite its importance, the state distribution in multi-site adsorption remains largely unknown in most practical systems. Experimentally, standard characterization tools such as ellipsometry \cite{Chen2025simultaneous}, quartz crystal microbalance with dissipation monitoring (QCM-D) \cite{Scott2008protein, Jordan2008qcmd}, and surface plasmon resonance \cite{HojjatJodaylami2025surface} measure total adsorbed mass or layer thickness but cannot resolve the conformational differences between one-, two-, and more-point attached chains, as these species differ only in their internal loop-tail statistics rather than in gross layer properties. Molecular dynamics simulations provide atomistic detail but are limited to microseconds or milliseconds \cite{Arkin2017polymer, Shen2024adsorption, Zhou2025adsorption, Dionne2006adsorption, Li2023long, Mao2024onepot, Rabe2011understanding, Shi2013selfassembly, Feng2022finding}. The evolution of multi-site adsorption state distributions, however, occurs over seconds to minutes due to high free-energy barriers arising from crowding and chain entanglements, rendering direct MD simulations computationally prohibitive \cite{Frantz1991kinetics, Rossner2017uniform, Duan2023sitespecific}.

Theoretical treatments also face difficulties. Existing kinetic models either assume a single adsorbed species (single-site approximation) or require unknown rate constants for every possible state transition \cite{Hasegawa1997adsorption, Zwanzig2001nonequilibrium, Tzlil2005flexible, Li2023long, Mao2024onepot, Rabe2011understanding, DiLeva2026polymer}. The problem is intrinsically coupled: the rate of transitioning from two-point to three-point attachment depends on the surrounding adsorbed layer composition, which itself evolves with time, creating a self-consistent but intractable system. Consequently, the distribution has remained unresolved by experimental, computational, or theoretical approaches, as it involves both fast intra-chain conformational rearrangements and slow surface adsorption-desorption kinetics, which operate on vastly different timescales \cite{Hasegawa1997adsorption, Lu2019nonsacrificial, Zhang2023enhancing}. The interfacial state distribution has thus remained a blind spot: invisible to experiments, unattainable by simulations, and intractable by theory. 

Here, we overcome this limitation by developing the State-Resolved Integral of First-Passage Times (SR-IFS) method, a new theoretical framework that directly addresses this long-standing challenge. The key innovation is a timescale decoupling strategy: we separate the fast internal adjustment of adsorbed polymers (intra-layer relaxation) from the slow external adsorption-desorption processes. This decoupling allows us to construct a coupled master equation in which all transition rates between different adsorption states are computed from constrained simulations and then used to iteratively solve for the self-consistent state distribution at each adsorption density. The SR-IFS method predicts not just the total adsorbed amount, but also the full state-resolved distribution ($p_1$, $p_2$, $p_3$) as a function of time, a quantity that is not accessible through alternative approaches \cite{Scott2008protein, Rossner2017uniform, Duan2023sitespecific}.

We demonstrate the feasibility of SR-IFS using 3-arm star-like polymers as a model system. We first validate the timescale separation by showing that internal relaxation times ($\tau_{\text{adj}}$) are consistently orders of magnitude shorter than adsorption timescales ($\tau_{\text{ads}}$), justifying our decoupling ansatz. We then apply SR-IFS to track the full temporal evolution of the interfacial state distribution, revealing a transition from three-point-dominated surfaces at low coverage to mixed states with significant two- and one-point fractions at later times. Finally, we show that tuning monomer binding energy provides systematic control over the final equilibrium state distribution, offering a design principle for tailored polymer interfaces. Our results establish SR-IFS as a generalizable framework that bridges the gap between microscopic kinetics and macroscopic interfacial properties. By providing access to the previously invisible state distribution of multi-site adsorbed polymers, this work lays a foundation for the rational design of polymer-functionalized nanomaterials \cite{Awasthi2020polymer, Sze2017single, Vreeker2025nanoporefunctionalized, Wang2024chronopotentiometric, Yusko2017realtime, Gal2016macromolecular, Hosono2020metalorganic, Imaoka2013macromolecular} whose performance hinges on detailed interfacial architecture.

The paper is organized as follows. Section Method presents the SR-IFS methodology for multi-site adsorption and describes the simulation setup. Section Result and Discussion applies the method to 3-arm star-like polymers, characterizing the evolving interfacial state distribution and the effect of binding energy. Section Conclusion concludes with a summary and outlook.

\begin{figure}
	\includegraphics[width=1\columnwidth]{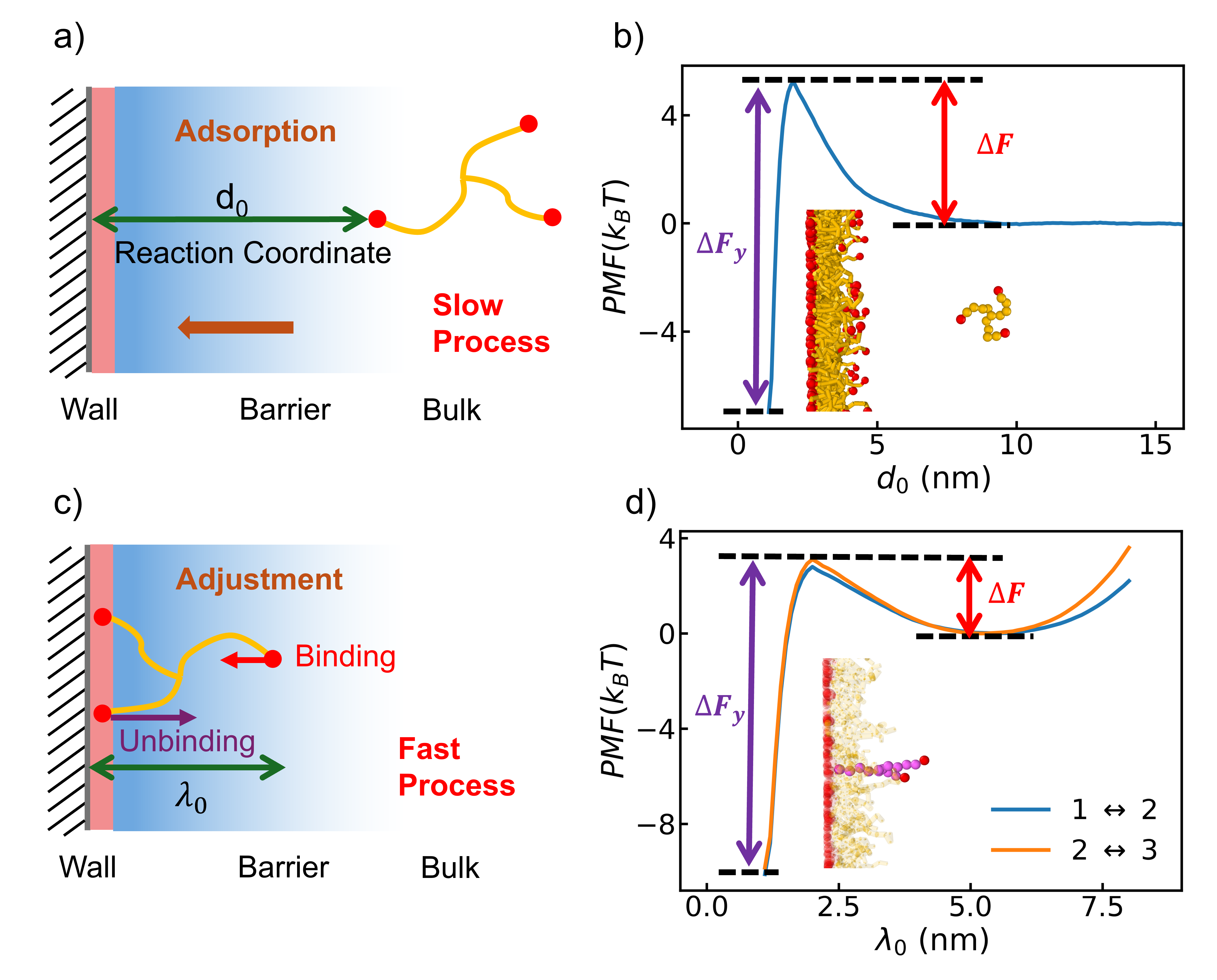}
	\caption{ (a) Schematic illustration of the diffusion process and (c) the adjustment process of a star-like polymer with three binding sites. The blue region indicates the energy barrier generated by adsorbed chains. In (a), the reaction coordinate $d_0$ for diffusion is defined as the distance between the wall and the binding monomer (red sphere). In (c), the reaction coordinate $\lambda_0$ for the adjustment process is defined as the distance between the wall and the free binding monomer (red sphere). Diffusion is a slow process, while adjustment is fast. (b-d) Potential of mean force (PMF) experienced by (b) a binding monomer of a free polymer at different distances $d_0$ and (d) a free binding monomer of an adsorbed polymer at different distances $\lambda_0$. The red vertical arrow indicates the binding energy barrier $\Delta F$, while the purple vertical arrows denote the unbinding energy barrier $\Delta F_y$. }\label{fig1}
\end{figure}

\begin{figure}
	\includegraphics[width= 1\columnwidth]{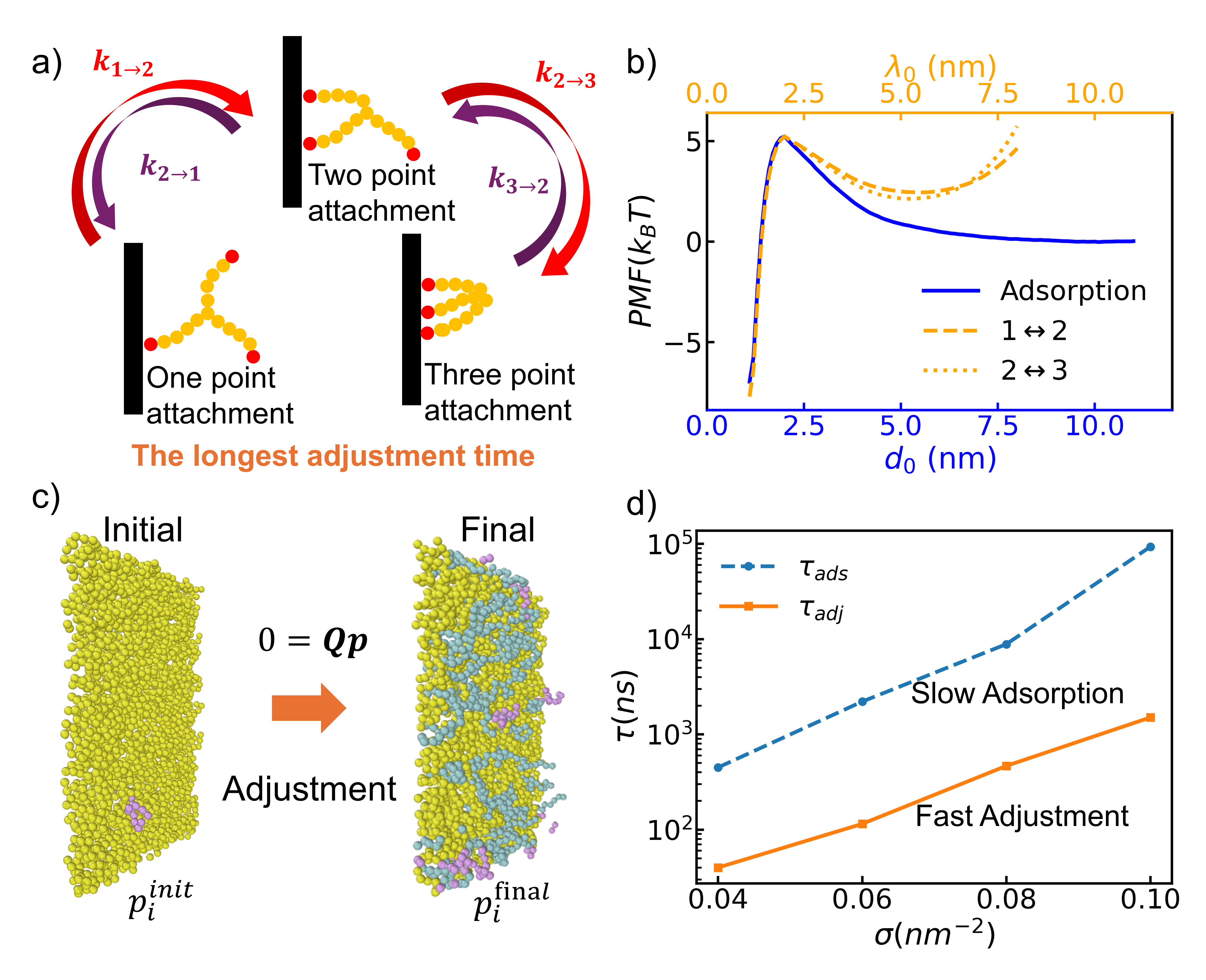}
	\caption{ (a) Illustration of the mutual adjustment among the one-, two-, and three-point adsorption states. The slowest adjustment time is taken as $\tau_{adj}$. (b) Potential of mean force for a monomer adsorbing from the bulk solution (blue) and for a monomer belonging to an already adsorbed chain (orange), including transitions among the three states. The adsorption density is $\sigma = 0.06$ nm$^{-2}$ in both cases. (c) Schematic of the adjustment process from an initial state distribution $p_i^{\text{init}}$ to the final distribution $p_i^{\text{final}}$. The one-, two-, and three-point attachments are shown in yellow, blue, and purple, respectively. (d) Adjustment time $\tau_{adj}$ and diffusion time $\tau_{ads}$ as functions of adsorption density $\sigma$. The binding energy of the binding monomers is set to $15 k_BT$. }\label{fig2}
\end{figure}

\section{Method}

\subsection{First-Passage Time Framework for Polymer Adsorption}

We begin by recalling the essential elements of the first-passage time framework that form the basis of our state-resolved extension \cite{Zhang2024barrierlimited, Tang2020prediction, Tang2024predicting, Huang2024heterogeneous, Huang2025extension}. The adsorption of a free polymer chain onto a planar surface can be mapped to the diffusion of a binding monomer across a free energy barrier generated by pre-adsorbed chains and surrounding monomers (Fig.~\ref{fig1}(a)). The characteristic timescale for this activated process is the first passage time, given by the Kramers-type expression \cite{Zwanzig2001nonequilibrium}:
\begin{equation}
	\tau_{\text{bind}}(\Delta F) = \frac{d_{s}^{2}}{D_{m}} \frac{\pi}{2\Delta F} \exp(\Delta F),
	\label{eq:tau_ads}
\end{equation}
where $\Delta F$ is the dimensionless free energy barrier, $d_s$ is the barrier width, and $D_m$ is the monomer diffusion coefficient. Similarly, the unbinding process is characterized by its own first passage time, $\tau_{\text{unbind}}(\Delta F_y)$, where $\Delta F_y$ is the desorption barrier. Both $\Delta F$ and $\Delta F_y$, along with the barrier widths $d_s$ and $d_y$, can be extracted from the potential of mean force (PMF) profiles obtained via constrained molecular simulations (Fig.~\ref{fig1}(b,d)). The PMF calculation details are provided in Section~\nameref{sec:simulation_details}. These first passage times serve as the input parameters for the state-resolved kinetic model developed below.

\subsection{State-Resolved IFS for Multi-Site Adsorption}

\subsubsection{Coupled Dynamics of Multi-State Adsorption}

We illustrate the SR-IFS method using 3-arm star-like polymers as a model system. In this case, the adsorbed layer consists of chains attached via one, two, or three anchor points. The evolution of the adsorption densities $\sigma_1$, $\sigma_2$, and $\sigma_3$ for these three states (Fig.~\ref{fig2}(a)) is governed by a system of coupled differential equations:
\begin{equation}
	\begin{dcases}
	\frac{d \sigma_1}{dt} = \frac{3 c_p d_s}{2 \tau_{bind}} + 2k_{2-1}\sigma_2 -2k_{1-2}\sigma_1 - k_{unbind} \sigma_1 \\
	\frac{d \sigma_2}{dt} = 3k_{3-2} \sigma_3 + 2k_{1-2} \sigma_1
	- 2k_{2-1} \sigma_2 - k_{2-3} \sigma_2 \\
	\frac{d \sigma_3}{dt} = k_{2-3} \sigma_2 - 3k_{3-2} \sigma_3	
	\end{dcases}
	\label{eq:adsorption_equation_system}
\end{equation}
where $c_p$ is the bulk polymer concentration, $k_{i-j}$ are the rate constants for transitions between adsorption states (Fig.~\ref{fig1}(c)), and the numerical coefficients (3, 2, 1) account for the number of binding monomers involved in each process. The rate constants are related to the corresponding first passage times via $k_{i-j} = 1/2\tau_{i-j}$. The adsorption rate from the bulk, $3 c_p d_s / 2 \tau_{bind}$, describes free polymers attaching to the surface via a single monomer, while desorption occurs exclusively from the one-point attached state.

The coupled system in Eq.~\eqref{eq:adsorption_equation_system} cannot be solved directly because the transition rates $k_{i-j}$ and the binding-unbinding first passage times $\tau_{bind}$ and $\tau_{unbind}$ depend on the unknown composition of the adsorbed layer. Specifically, the state distribution ($p_1$, $p_2$, $p_3$), where $p_i = \sigma_i / \sigma$ and $\sigma = \sigma_1 + \sigma_2 + \sigma_3$ is the total adsorption density. This composition dependence arises because the energy barriers experienced by a polymer depend on the steric environment created by all adsorbed chains, which varies with the relative fractions of one-, two-, and three-point attachments. A method to self-consistently determine this state distribution is therefore required.

\subsubsection{Timescale Decoupling and Self-Consistent State Distribution}

The central physical insight underlying our approach is that the internal rearrangement of adsorbed polymers, \ie, transitions between one-point, two-point, and three-point attachments, occurs on a much faster timescale than the overall adsorption and desorption of chains from the bulk. This separation is supported by the PMF comparison in Fig.~\ref{fig2}(b), where the energy barriers for adjustment are consistently lower than those for adsorption from the bulk. We therefore invoke a timescale decoupling: the adsorbed layer relaxes to a quasi-equilibrium state distribution instantaneously on the timescale of changes in total adsorption density.

This decoupling separates the full dynamics into two components. The slow component governs the evolution of the total adsorption density $\sigma$:
\begin{equation}
	\frac{d\sigma}{dt} = \frac{3 c_p d_s}{2 \tau_{bind}} -  \frac {p_1 \sigma}{2 \tau_{unbind}},
	\label{eq:adsorption_kinetics}
\end{equation}
where $\tau_{bind}$ and $\tau_{unbind}$ depend on the instantaneous state distribution ($p_1$, $p_2$, $p_3$). The fast component describes the internal relaxation, which reaches steady state at each density:
\begin{equation}
	\begin{dcases}
	0 = 2k_{2-1} p_2 -2k_{1-2} p_1 \\
	0 = 3k_{3-2} p_3 + 2k_{1-2} p_1 - 2k_{2-1} p_2 - k_{2-3} p_2 \\
	0 = k_{2-3} p_2 - 3k_{3-2} p_3 
	\end{dcases}
	\label{eq:relaxation}
\end{equation}
with the normalization $\sum p_i = 1$.

To determine the state distribution ($p_1$, $p_2$, $p_3$) at a given total adsorption density $\sigma$, we employ an iterative, self-consistent procedure illustrated in Fig.~\ref{fig2}(c). An initial state distribution $p_i^{\text{init}}$ is first guessed. A simulation system with the corresponding adsorbed layer composition at density $\sigma$ is then constructed, and constrained simulations are performed to extract the PMF profiles for both bulk adsorption and internal transitions (Fig.~\ref{fig2}(b)). From these PMF profiles, the first passage times $\tau_{bind}$, $\tau_{unbind}$, and $\tau_{i-j}$ are obtained, and the rate constants $k_{i-j}$ are computed. Substituting $k_{i-j}$ into Eq.~\eqref{eq:relaxation} yields the equilibrium distribution $p_i^{\text{equ}}$, which is compared with $p_i^{\text{init}}$. If convergence is achieved, the correct state distribution $p_i^{\text{final}}$ has been identified. Otherwise, $p_i^{\text{init}}$ is updated and the procedure is repeated.

This procedure yields the self-consistent state distribution $p_i(\sigma)$ as a function of total adsorption density. With $p_i(\sigma)$ known, $\tau_{bind}(\sigma)$ and $\tau_{unbind}(\sigma)$ are fully determined, and Eq.~\eqref{eq:adsorption_kinetics} can be numerically integrated to obtain the time evolution of the total adsorption density $\sigma(t)$ and, consequently, the time-dependent state distribution $p_i(t)$.

\subsubsection{Validation of Timescale Separation}

To justify the timescale decoupling, we quantify the internal relaxation time $\tau_{adj}$ and compare it with the characteristic adsorption timescale $\tau_{ads}$. The relaxation dynamics of the state distribution is governed by the coefficient matrix $\mathbf{Q}$ of Eq.~\eqref{eq:relaxation}:
\begin{equation}
	\mathbf{Q} = \begin{bmatrix}
		-2k_{1-2} &2k_{2-1} &0 \\
		2k_{1-2} &-(2k_{2-1} + k_{2-3}) &3k_{3-2}\\
		0 &k_{2-3} &-3k_{3-2}
		\label{eq:coefficient matrix}
	\end{bmatrix}.
\end{equation}
This system is a continuous-time Markov process \cite{Puterman2014markov} with eigenvalues $\lambda_1=0$, $\lambda_2$, and $\lambda_3$. The internal relaxation time is determined by the slowest non-zero mode: $\tau_{adj} = 1/|\lambda_2|$. The adsorption timescale is defined as:
\begin{equation}
	\tau_{ads} = \frac{\Delta \sigma}{\left|\frac{3 c_p d_s}{2 \tau_{bind}} - \frac{p_1 \sigma}{2 \tau_{unbind}}\right|},
\end{equation}
where $\Delta \sigma = 0.02$ nm$^{-2}$ is the discretization interval used in the numerical integration.

Figure~\ref{fig2}(d) compares $\tau_{adj}$ and $\tau_{ads}$ across the density range of interest. For all densities, $\tau_{ads}$ exceeds $\tau_{adj}$ by more than an order of magnitude, with the difference growing at higher densities. This clear separation of timescales validates our decoupling assumption and establishes the physical foundation of the SR-IFS method.

\subsection{Simulation Details and Potential of Mean Force Calculation}
\label{sec:simulation_details}

The first passage times $\tau_{bind}$, $\tau_{unbind}$, and $\tau_{i-j}$ are computed from the energy barrier parameters extracted from the potential of mean force (PMF) profiles, as shown in Fig.~\ref{fig1}(b,d). In this work, we use a coarse-grained model to calculate these PMF profiles.

The non-bonded interactions between polymer monomers are modeled using a truncated and shifted Lennard-Jones (LJ) potential \cite{Grest1986molecular}: 
\begin{equation}
	U_{\mathrm{LJ}}(r)=
	\begin{cases}
		4 \varepsilon\left[\left(\dfrac{\sigma_{\rm m}}{r}\right)^{12}-\left(\dfrac{\sigma_{\rm m}}{r}\right)^{6}\right]-U_{\mathrm{cut}}, & r \leqslant r_{\mathrm{cut}} \\
		0, & r>r_{\mathrm{cut}}
	\end{cases}\label{eq:LJ}
\end{equation}
where $\sigma_m = 1.0$ nm is the diameter of a coarse-grained monomer, $\varepsilon$ sets the energy scale ($k_{\mathrm{B}}T$), and $r_{\text{cut}} = 2^{1/6}\sigma_m$ with $U_{\text{cut}}= 4 \varepsilon[(\sigma_m/r_{cut})^{12}-(\sigma_m/r_{\rm cut})^6]$ ensuring continuity at $r_{\text{cut}}$, modeling a good solvent.

Adjacent monomers are connected by a FENE potential \cite{Grest1986molecular}:
\begin{equation}
	U_b(r) =
	\begin{cases}
		-\frac{1}{2}kl_{\rm max}^2 \ln\left[1-\left(r/l_{\rm max} \right)^2\right], &r \leqslant l_{\rm max} \\
		\infty, &r > l_{max} 
	\end{cases}
	\label{eq:FENE}
\end{equation}
with $l_{\text{max}} = 1.5\sigma_m$ and $k = 30.0\varepsilon/\sigma_{m}^{2}$ to prevent chain crossing.

The attractive interaction between the planar wall and any binding monomer is also modeled using a LJ potential. The well-depth of this potential is set as $\varepsilon=15k_BT$, where $k_B$ is the Boltzmann constant and $T$ is the temperature. This choice of well-depth is grounded in our previous studies on star-like polymer adsorption toward planar surfaces \cite{Tang2020prediction}, which generated a binding energy of 36 kJ/mol (or 14.52 $k_BT$) between the arm-ends of stars and the Au-NP surface. The cut-off radius for the LJ potential is defined as $r_{cut}=2.0 \sigma_m$, where $\sigma_m$ represents the characteristic length scale related to the monomer. During the simulation, the temperature was controlled by the Nos\'e--Hoover thermostat.

Length and time units are set to nanometers (nm) and nanoseconds (ns), respectively. One coarse-grained monomer has size $\sigma_m = 1.0$ nm. The time unit $\tau = 150$ ns is defined based on mapping the self-diffusion coefficient of a 40-monomer chain to experimental values \cite{Zettl2009selfdiffusion}. The simulation box is fixed at $100 \mathrm{nm} \times 40 \mathrm{nm} \times 40 \mathrm{nm}$. The bulk concentration of polymers ($c_p$) is $4.8 \times 10^{-4}$ nm$^{-3}$ for all polymer types considered in this paper. The arm number is fixed to three for all polymers. The diffusion coefficients of polymers, calculated from mean squared displacement, are $D_m = 1.5$ nm$^2$/ns (for monomers) and $D_p = 0.2$ nm$^2$/ns.

The Potential of Mean Force (PMF) along the reaction coordinate $d_0$ (distance from wall to polymer center of mass) is calculated using the Adaptive Biasing Force (ABF) method \cite{Fiorin2013using, Thompson2022lammps, Comer2015adaptive} as implemented in the Colvars module \cite{Fiorin2013using} of LAMMPS \cite{Thompson2022lammps}. The ABF algorithm \cite{Comer2015adaptive} applies a bias force $-F(r_0)$ opposite to the running average of the instantaneous force $F$ at a given position $r_0$. This actively flattens the free energy landscape along $d_0$ during the simulation, thereby enhancing sampling across the entire reaction coordinate in a single, continuous process. This on-the-fly capability makes ABF well suited for studying dynamic processes like polymer adsorption over a wide range of the reaction coordinate. The PMF is obtained directly by integrating the negative of the average force, which is concurrently estimated as the simulation proceeds. The PMF at a given point $d_0$ is given by: 
\begin{equation}
	U_{\text{PMF}}(d_0) = \int_{\infty}^{d_0} -F(r) dr.
	\label{eq:PMF}
\end{equation}
The calculated PMF profiles provide the parameters ($U_{\text{max}}$, $U_{\text{min}}$, $d_s$, $d_y$, etc.) essential for computing the first passage times $\tau_{\text{bind}}$, $\tau_{\text{unbind}}$, and $\tau_{i-j}$ in Eq.~\eqref{eq:tau_ads} and Eq.~\eqref{eq:relaxation}.

\section{Results and Discussion}

\subsection{Energy Barriers and Timescale Separation: Validation of the SR-IFS Assumption}

\begin{figure}
	\includegraphics[width= 1\columnwidth]{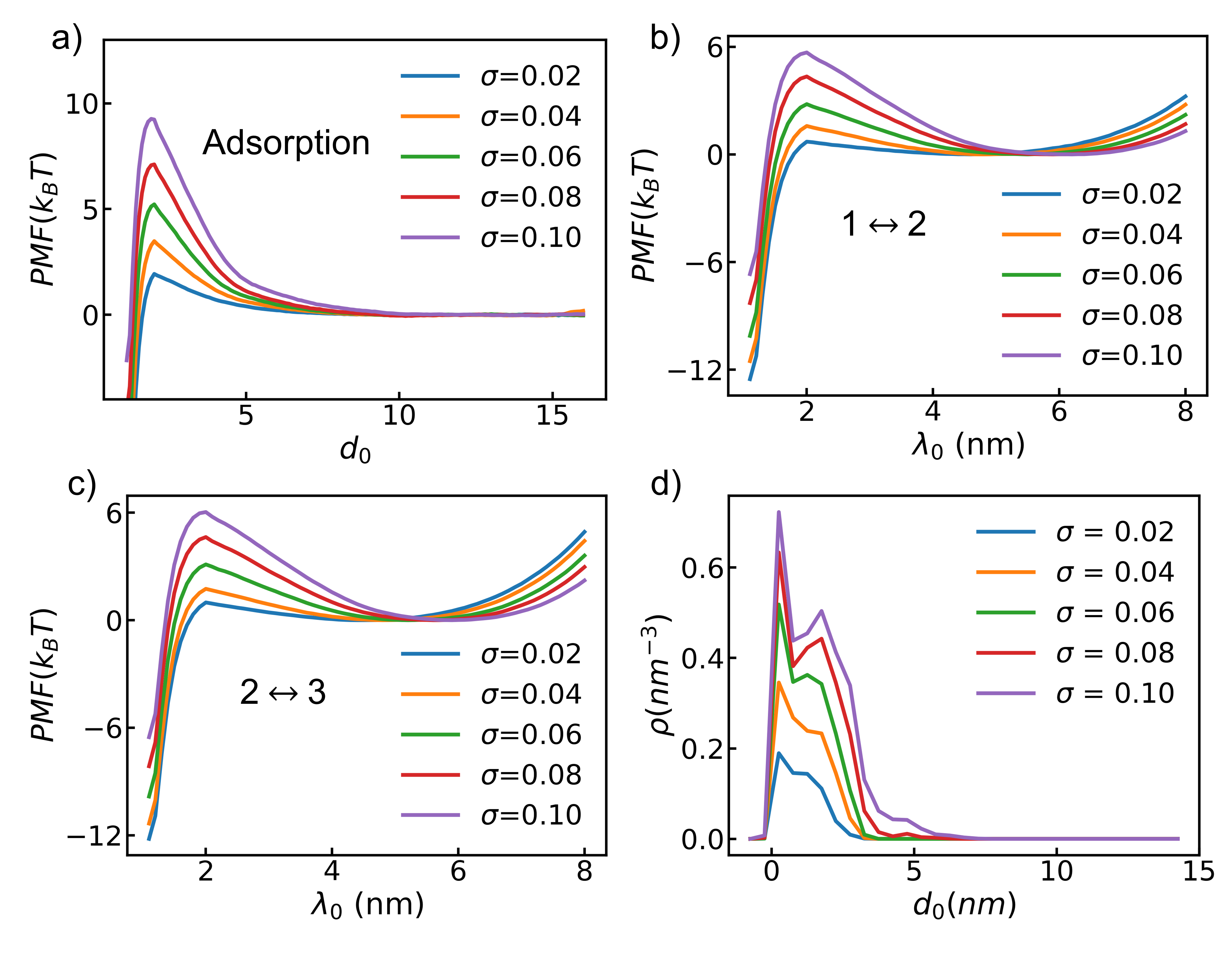}
	\caption{ (a) Potential of mean force experienced by a monomer of a free polymer chain in bulk solution approaching a surface covered with adsorbed chains at adsorption densities $\sigma = 0.02, 0.04, 0.06, 0.08$, and $0.10$ nm$^{-2}$. (b-c) Potential of mean force experienced by a free binding monomer of an adsorbed polymer under (b) one-point and (c) two-point attachment at the same adsorption densities. (d) Average monomer density distributions for adsorbed surfaces at $\sigma = 0.02, 0.04, 0.06, 0.08$, and $0.10$ nm$^{-2}$.}\label{fig3}
\end{figure}

The central assumption of the SR-IFS method is that internal rearrangement of adsorbed polymers occurs on a significantly faster timescale than the overall adsorption and desorption kinetics. To test this assumption, we first extract the energy barriers governing each process from potential of mean force (PMF) calculations, then compute the corresponding first passage times.

The PMF experienced by a binding monomer approaching the surface from the bulk is shown in Fig.~\ref{fig3}(a) for several total adsorption densities $\sigma$. A free energy barrier $\Delta F$ is observed in front of the surface, increasing monotonically with $\sigma$ from approximately 2 $k_BT$ at $\sigma = 0.02$ nm$^{-2}$ to over 9 $k_BT$ at $\sigma = 0.10$ nm$^{-2}$. This barrier arises from steric hindrance created by pre-adsorbed chains and intensifies as the surface becomes more crowded. For the adjustment processes, Figs.~\ref{fig3}(b) and (c) show the PMF barriers for transitions from one-point to two-point ($\Delta F_{1\rightarrow2}$) and from two-point to three-point ($\Delta F_{2\rightarrow3}$) attachments, respectively. These barriers exhibit a similar increasing trend with $\sigma$, though with consistently lower magnitudes than the bulk adsorption barrier. The density profiles in Fig.~\ref{fig3}(d) confirm that the growing steric hindrance arises from enhanced packing of adsorbed chains: as $\sigma$ increases, the monomer density near the surface becomes sharper and more intense, and for $\sigma > 0.08$ nm$^{-2}$, significant density appears at distances beyond 5 nm, reflecting the emergence of loops and tails from one- and two-point attached chains.

\begin{figure}
	\includegraphics[width= 1\columnwidth]{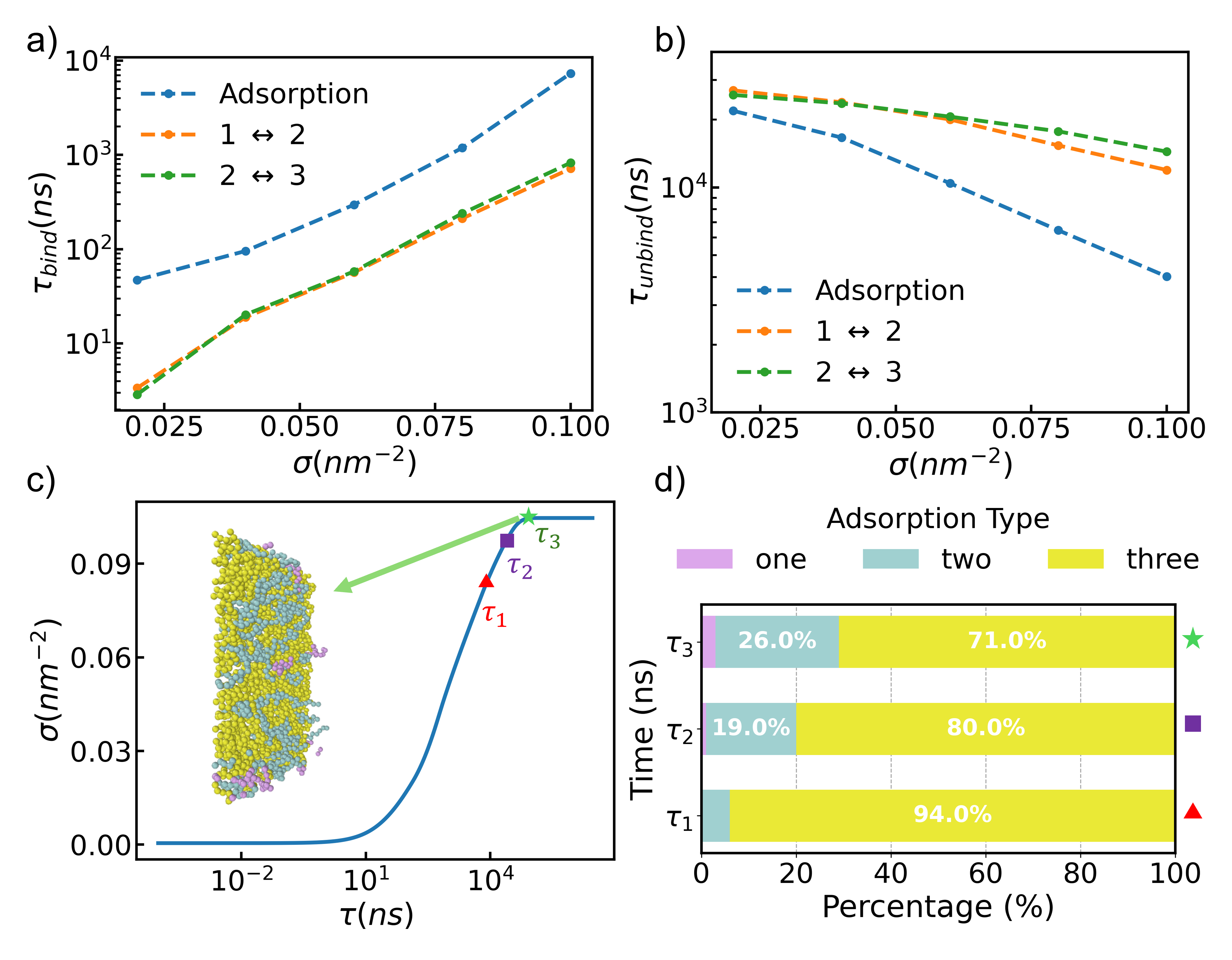}
	\caption{ (a-b) First passage times for (a) binding and (b) unbinding to the planar surface as functions of adsorption density $\sigma$, comparing the reference adsorption process, the one-to-two-point adjustment, and the two-to-three-point adjustment. (c) IFS-predicted adsorption kinetics of polymers onto a planar surface. Inset: snapshot of adsorbed polymers on the surface, with one-, two-, and three-point attached polymers shown in purple, blue, and yellow, respectively. (d) Adsorption state distributions at three representative times: $\tau_1 = 6.2 \times 10^3$ ns, $\tau_2 = 3.2 \times 10^4$ ns, and $\tau_3 = 1.3 \times 10^5$ ns.}\label{fig4}
\end{figure}

The first passage times are computed from the PMF barriers using Eq.~\eqref{eq:tau_ads}. Figure~\ref{fig4}(a) presents $\tau_{bind}$ as a function of $\sigma$ for bulk adsorption and for the internal adjustment processes ($1\rightarrow2$ and $2\rightarrow3$). For bulk adsorption, $\tau_{bind}$ increases by approximately two orders of magnitude, \ie, from 47 to 7318 ns, as $\sigma$ increases from 0.02 to 0.10 nm$^{-2}$. The adjustment processes show the same trend, with $\tau_{bind}$ rising from 3 to 820 ns. Across the entire density range, the bulk adsorption timescale consistently exceeds the adjustment timescales by approximately one order of magnitude. Figure~\ref{fig4}(b) shows the corresponding unbinding first passage times $\tau_{unbind}$. In contrast to $\tau_{bind}$, $\tau_{unbind}$ decreases with $\sigma$ for all processes due to the enhanced excluded volume that lowers the desorption barrier. For bulk adsorption, $\tau_{unbind}$ drops from 22,000 to 4,000 ns, while the adjustment processes show a more modest decrease from 2,600 to 1,440 ns.

A more quantitative measure of the timescale separation is obtained by computing the internal relaxation time $\tau_{adj}$ from the eigenvalues of the transition matrix $\mathbf{Q}$ (see Eq.~\ref{eq:coefficient matrix}) and comparing it with the characteristic adsorption timescale $\tau_{ads}$. As shown in Fig.~\ref{fig2}(d), $\tau_{ads}$ exceeds $\tau_{adj}$ by more than an order of magnitude across the entire density range (0.04--0.10 nm$^{-2}$), with the gap widening at higher densities. This separation validates the timescale decoupling assumption that underpins the SR-IFS method and confirms that the adsorbed layer reaches a quasi-equilibrium state distribution on the timescale of changes in total adsorption density.

\subsection{State-Resolved Prediction of Interfacial Distributions}

Having validated the timescale decoupling, we now apply the SR-IFS method to predict the total adsorption kinetics and, uniquely, the evolution of the interfacial state distribution ($p_1$, $p_2$, $p_3$) for 3-arm star-like polymers adsorbing from bulk solution ($c_p = 4.8 \times 10^{-4}$ nm$^{-3}$) onto a planar surface.

Figure~\ref{fig4}(c) shows the total adsorption density $\sigma(t)$ obtained by integrating Eq.~\eqref{eq:adsorption_kinetics}. The adsorption proceeds through three distinct regimes: a slow initial increase ($t < 10$ ns), a rapid growth phase ($10 < t < 10^5$ ns), and saturation at $\sigma \approx 0.11$ nm$^{-2}$ for $t > 10^5$ ns. The saturation corresponds to a dynamical steady state where the adsorption and desorption fluxes balance. This kinetic profile is consistent with the expected behavior of polymer adsorption with high barriers, and is comparable with our previous experimental and computational results for similar systems \cite{Tang2020prediction, Tang2024predicting, Huang2024heterogeneous, Zhang2024barrierlimited, Huang2025extension}.

The SR-IFS method provides not only the total adsorption density but also the time-dependent state distribution $p_i(t)$, a quantity that is experimentally inaccessible and beyond the reach of conventional simulations. Figure~\ref{fig4}(d) presents the distributions at three representative times (marked by triangle, square, and star in Fig.~\ref{fig4}(c)): $\tau_1 = 6.2 \times 10^3$ ns, $\tau_2 = 3.2 \times 10^4$ ns, and $\tau_3 = 1.3 \times 10^5$ ns. At early times ($\tau_1$), the interface is dominated by three-point attached chains ($p_3 \approx 1$). By $\tau_2$, a significant fraction of two-point attachments has emerged ($p_2 \approx 0.19$), while $p_3$ decreases correspondingly. At the equilibrium state ($\tau_3$), the interface consists of a mixed distribution: approximately 71\% three-point, 26\% two-point, and 3\% one-point attachments. The inset in Fig.~\ref{fig4}(c) provides a representative snapshot of this equilibrium interface, where three distinct attachment states are visible.

Even at equilibrium, a non-negligible fraction ($\sim 30\%$) of chains are not fully anchored by all three arms. This coexistence of one-, two-, and three-point attachments reflects the balance between the configurational entropy gain of releasing arms from the surface and the energetic cost of desorption, a balance that the SR-IFS method resolves self-consistently. While experiments such as ellipsometry and QCM-D provide ensemble-averaged measures of adsorbed mass and layer thickness \cite{Chen2025simultaneous, Scott2008protein, Jordan2008qcmd, HojjatJodaylami2025surface}, and simulations offer detailed trajectories at short timescales \cite{Frantz1991kinetics, Rossner2017uniform, Duan2023sitespecific}, the state-resolved information obtained here complements these approaches by providing access to the distribution of attachment states that is not directly accessible from either technique alone.

\subsection{Tuning Interfacial State Distributions via Binding Energy}

\begin{figure}
	\includegraphics[width= 1\columnwidth]{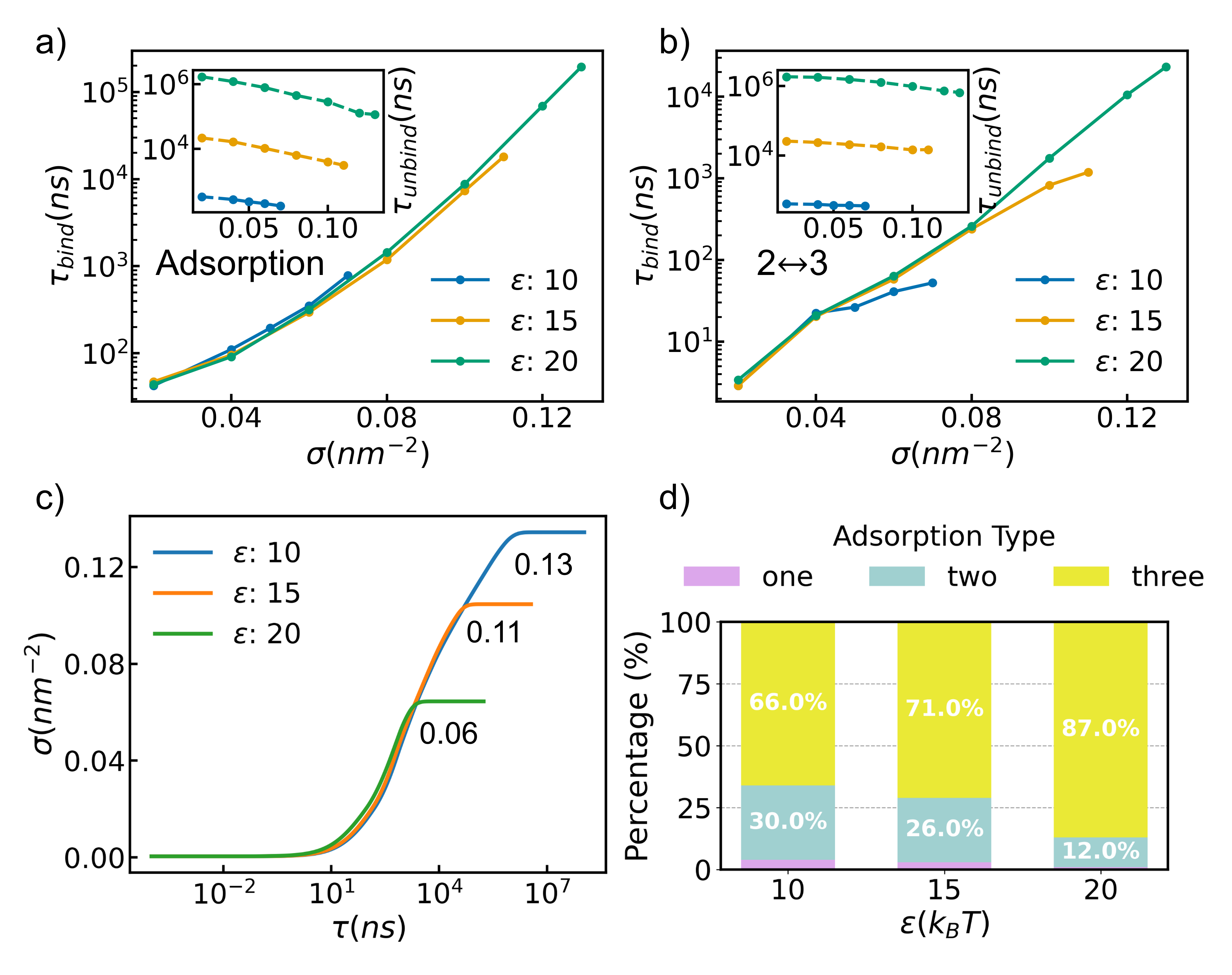}
	\caption{ (a) First passage time for binding to the planar surface as a function of adsorption density $\sigma$ at binding energies $\varepsilon = 10, 15$, and $20$ $k_BT$. Inset: first passage time for unbinding. (b) First passage time for binding for the $2 \leftrightarrow 3$ adjustment as a function of $\sigma$. Inset: first passage time for unbinding. (c) Adsorption kinetics predicted by the IFS method at $\varepsilon = 10, 15$, and $20$ $k_BT$. The concentration of binding monomers is $c_p = 4.8 \times 10^{-4}$ nm$^{-3}$. (d) Final equilibrium state distribution for different binding energies. }\label{fig5}
\end{figure}

The binding energy $\varepsilon$ between monomers and the planar surface is a key experimental parameter for controlling polymer adsorption \cite{Tang2020prediction, Tang2024predicting, Thoms2024simple}. We use SR-IFS to investigate how $\varepsilon$ influences both the adsorption kinetics and the final equilibrium state distribution.

Figure~\ref{fig5}(a) shows $\tau_{bind}$ for the bulk adsorption process at binding energies $\varepsilon = 10, 15,$ and $20$ $k_BT$. At a given adsorption density $\sigma$, $\tau_{bind}$ is essentially independent of $\varepsilon$, increasing exponentially with $\sigma$ in all three cases. This indicates that the binding energy does not affect the barrier for a free polymer approaching an already formed adsorbed layer; the barrier is dominated by steric repulsion from the pre-adsorbed chains rather than by the monomer--surface attraction. In contrast, the unbinding timescale $\tau_{unbind}$ (inset of Fig.~\ref{fig5}(a)) is highly sensitive to $\varepsilon$: higher binding energies result in longer $\tau_{unbind}$ at all densities, as expected from the deeper potential well that traps adsorbed monomers.

The adjustment processes exhibit a qualitatively similar $\varepsilon$-dependence to the bulk adsorption process. As shown in Fig.~\ref{fig5}(b), $\tau_{bind}$ for the $2\rightarrow3$ transition increases exponentially with $\sigma$ and is largely independent of $\varepsilon$, with a weaker but noticeable dependence at higher densities. The overall magnitude of $\tau_{bind}$ is approximately one order of magnitude lower than that for bulk adsorption (Fig.~\ref{fig5}(a)), reflecting the lower barrier associated with rearranging an already adsorbed chain. The $\tau_{unbind}$ for the $2\leftrightarrow3$ adjustment (inset of Fig.~\ref{fig5}(b)) shows a similar decreasing trend with $\sigma$ as the bulk case, though with a smaller slope and a weaker $\varepsilon$-dependence, indicating that the adjustment processes are less sensitive to crowding.

These $\varepsilon$-dependent kinetics translate into macroscopic adsorption behavior, as shown in Fig.~\ref{fig5}(c). For $\varepsilon = 10$ $k_BT$, the adsorption saturates rapidly at $\sigma \approx 0.06$ nm$^{-2}$ within $10^3$ ns. Increasing $\varepsilon$ to 15 and 20 $k_BT$ progressively raises the equilibrium adsorption density to approximately 0.09 and 0.13 nm$^{-2}$, respectively, with saturation times extending beyond $10^6$ ns for the highest binding energy.

The SR-IFS method further reveals how $\varepsilon$ controls the final equilibrium state distribution (Fig.~\ref{fig5}(d)). As $\varepsilon$ increases from 10 to 20 $k_BT$, the fraction of three-point attachments in the equilibrium interface rises from 66\% to 87\%. Concurrently, the two-point fraction decreases from 30\% to 12\%, and the one-point fraction drops from 4\% to 1\%. Stronger binding energies thus favor fully anchored chains, yielding a more uniformly attached and mechanically robust interface; weaker binding energies produce interfaces with significant fractions of partially attached chains, which may offer greater conformational flexibility. The ability to predict such state-resolved design rules is a unique feature of the SR-IFS method, providing direct guidance for tailoring polymer interfaces for specific applications.

\section{Conclusion}
We have developed the State-Resolved Integral of First-Passage Times (SR-IFS) method, which resolves for the first time the time-dependent distribution of attachment states ($p_1$, $p_2$, $p_3$) within multi-site adsorbed layers. The method decouples fast intra-layer adjustments from slow adsorption kinetics and determines the interfacial composition self-consistently at each adsorption density. Using 3-arm star-like polymers as a model system, we validated the timescale separation underpinning the method and found that the interfacial state distribution evolves from three-point dominance at early stages to a mixed state with significant one- and two-point fractions at equilibrium. The final distribution can be tuned by the monomer binding energy, with stronger binding favoring fully anchored chains. By resolving the state distribution, a quantity not directly accessible from ensemble-averaged experiments \cite{Chen2025simultaneous, Scott2008protein, Jordan2008qcmd, HojjatJodaylami2025surface} or short-timescale simulations \cite{Frantz1991kinetics, Rossner2017uniform, Duan2023sitespecific}, our SR-IFS method provides a framework for connecting microscopic kinetics to macroscopic interfacial properties and offers a predictive basis for the rational design of polymer-functionalized interfaces.

\section{Acknowledgments}

The financial support from the National Natural Science Foundation of China under Grant Nos. 12374207, 12347102, and 12174184, the Fundamental and Interdisciplinary Disciplines Breakthrough Plan of the Ministry of Education of China (JYB2025XDXM502), the Natural Science Foundation of Jiangsu Province(No. BK20233001), and the Innovation Program for Quantum Science and Technology (2024ZD0300101) are acknowledged. This research work is supported by the Big Data Computing Center of Southeast University, and the authors also thank the super computing resources at the Beijing Super Cloud Computing Center (BSCC).

\bibliography{bibtex}

\end{document}